\documentclass{article}
\usepackage{LaThuileFPSpro}
\usepackage{epsfig}
\begin{document}
\title{RECENT CHARM RESULTS FROM CLEO-C}
\author{
  Istv\'an Dank\'o\\
  (for the CLEO Collaboration)\\
  {\em Rensselaer Polytechnic Institute}\\
  {\em Troy, NY 12180, USA} \\
  }
\maketitle

\baselineskip=11.6pt

\begin{abstract}
The CLEO-c experiment has been collecting data at the charm-threshold region. A selection of recent results on charmed meson and charmonia decays are presented.
\end{abstract}
\newpage
\section{Introduction}
The CLEO-c experiment has been taking data at the CESR symmetric $e^+e^-$ collider at the charm threshold region since 2003. The main goal of the experiment is to perform high precision measurements of hadronic branching fractions, leptonic decay constants, and semileptonic form factors of charmed mesons, together with an extensive study of QCD spectroscopy in the charmonium sector in order to provide rigorous constraints on the strong interaction theory, especially Lattice QCD calculations. If the theoretical calculations survive these tests they can be used to provide much needed theoretical input to extract quark mixing (CKM) matrix elements (e.g. $V_{ub}$, $V_{td}$ and $V_{ts}$), which remain limited by complications caused by strong interaction dynamics.

The selected topics discussed here are the measurement of the absolute branching fraction of Cabibbo-favored hadronic $D^0$, $D^+$, and $D_s$ decays; measurement of the leptonic decays, $D^+_{(s)} \to \ell^+ \nu$, and decay constants $f_{D_{(s)}}$; measurement of the $D_0$ mass; and a study of three-body hadronic decays of $\chi_{cJ}$.

Charged and neutral D mesons are produced at the $\psi(3770)$ which predominantly decays to $D^+D^-$ and $D^0\bar{D^0}$ with a total cross section of about $7$ nb. $D_s$ mesons are created at around $E_{\rm cm} = 4170$ MeV, where their production is dominated by $e^+e^- \to D_s^{\star \pm}D_s^{\mp}$ with a cross section about $0.9$ nb \cite{Polling:DsScan}. The $1 ^3P_J$ ($J=0,1,2$) charmonium states are produced in radiative $\psi(2S)$ decays with a branching fraction of ~9\% to each.
The main advantage of the CLEO experiment compared to B factories and fixed target experiments is the very clean experimental environment with low multiplicity final states, which arises from running at or slightly above production thresholds. Background is further reduced in $e^+e^- \to \psi(3770) \to D\bar{D}$ and $e^+e^- \to D_s^{\star}\bar{D}_s$ data by fully reconstructing (tagging) one of the $D_{(s)}$ decaying into a hadronic final state.

\section{Absolute $D^0, D^+, D_s$ hadronic branching fractions}
\label{Sec:Dhadron}

Precise knowledge of the absolute hadronic branching fractions of the $D^0$, $D^+$, $D_s$ mesons is important because they are used to normalize the decays of other charmed mesons and $B_{(s)}$ mesons. 

CLEO measures the absolute branching fraction of three $D^0$, six $D^+$, and six $D^+_s$ Cabibbo-favored hadronic decays using single tag and double tag events following a technique pioneered by the MARK-III Collaboration \cite{MARK3:Tagging}. In single tag events only one of the $D_{(s)}$ or $\bar{D}_{(s)}$ is reconstructed in a specific final state, while in double tag events both $D_{(s)}$ and $\bar{D}_{(s)}$ mesons are reconstructed in one of the hadronic final states. The single and double tag yield can be expressed as $n_i = N_{DD} {\cal B}_i \epsilon_i$ and $n_{ij} = N_{DD} {\cal B}_i {\cal B}_j \epsilon_{ij}$, where $N_{DD}$ is the number of $D^0\bar{D^0}$, $D^+D^-$, or $D_s^+D_s^-$ events produced; ${\cal B}_i$ is the branching fraction of decay mode $i$; $\epsilon_i$ and $\epsilon_{ij}$ are the single and double tag efficiencies. Then the absolute branching fractions can be obtained from the double and single tag ratios and efficiencies as
\begin{equation}
  {\cal B}_i = \frac{n_{ij}}{n_j} \frac{\epsilon_j}{\epsilon_{ij}}.
  \label{eq:AbsoluteBF} 
\end{equation}
Since $\epsilon_{ij} \approx \epsilon_i \epsilon_j$, the branching fraction is nearly independent of the efficiency of the tagging mode, and many systematic uncertainties cancel in the ratio.

There is a difference in the kinematics of $D$ and $D_s$ mesons. The $D$ and $\bar{D}$ mesons produced in $e^+e^- \to \psi(3770) \to D\bar{D}$ process have the same well defined energy (and momentum) in the center of mass frame of the colliding $e^+e^-$ beams ($E_D = E_{\rm beam}$).
In contrast, a pair of $D_s$ mesons is produced in $e^+e^- \to D_s^{\star \pm}D_s^{\mp}$ followed by the decay $D_s^{\star \pm} \to \gamma D_s^{\pm}$ (96\%) or $\pi^0 D_s^{\pm}$ (4\%). Therefore, the $D_s$ produced directly has a well defined energy and momentum in the $e^+e^-$ center of mass frame, while the secondary $D_s$ from the $D_s^{\star}$ decay has a much broader momentum distribution around the same value. This difference in kinematics leads to a slightly different  selection strategy of $D\bar{D}$ and $D_s^{\pm}D_s^{\mp}$ events.

\begin{figure}[h]
  \centering
  \epsfig{file=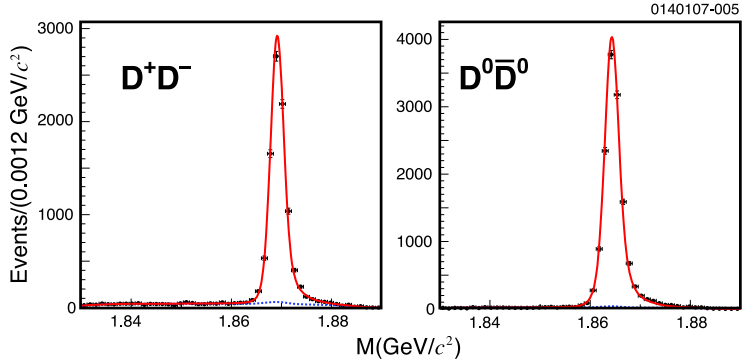,height=1.9in}
  \caption{\it
    Beam-constrained mass distribution of $D$ ($\bar{D}$) candidates in double tag events summed over all decay modes.
    \label{Fig:DDbarDoubleTags} }
\end{figure}

In order to identify (tag) the $D$ mesons, we use $\Delta E = E_{D} - E_{\rm beam}$ and the beam-constrained mass, $M_{\rm bc} = \sqrt{E_{\rm beam}^2 - (\vec{p}_{D})^2}$, where $E_D$ and $\vec{p}_D$ are the energy and three-momentum of the reconstructed $D$ meson candidate. Substituting the beam energy for $E_D$ improves the mass resolution of $D$ candidates by an order of magnitude, to about $2$ MeV/$c^2$. $\Delta E$ peaks around zero and $M_{\rm bc}$ peaks at the nominal $D$ mass.  We require $\Delta E$ to be consistent with zero within ~3 standard deviations, and extract the number of single and double tags from a fit to the one-dimensional and two-dimensional $M_{\rm bc}$ distributions, respectively. Fig. \ref{Fig:DDbarDoubleTags} illustrates the beam-constrained mass distribution for double tag events summed over all decay modes. In $281$ pb$^{-1}$ data, we reconstruct $230,000$ single tag and $13,575 \pm 120$ double tag $D^0 \bar{D}^0$ events, and $167,000$ single tag and $8,867 \pm 97$ double tag $D^+D^-$ events.

\begin{figure}
  \centering
  \epsfig{file=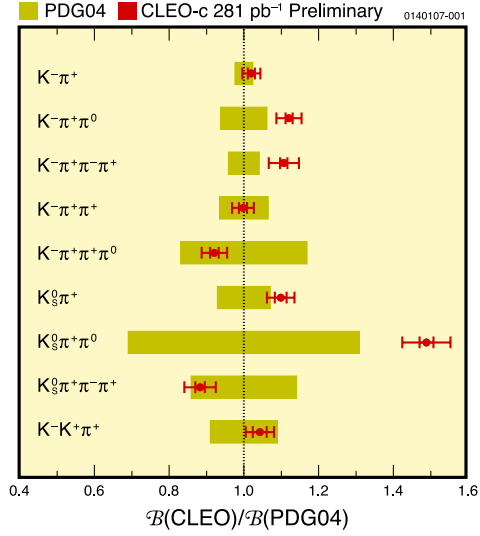,height=2.8in}
  \caption{\it
    Ratio of preliminary $D$ hadronic branching fractions to the 2004 PDG averages (dots). The shaded bars represent the errors in the PDG averages.
    \label{Fig:DBFvsPDG} }
\end{figure}

The $D^0$ and $D^+$ branching fractions are determined from a simultaneous least squares ($\chi^2$) fit to all $D^0$ and $D^+$ single and double tag yields. The fit properly takes into account correlations among all statistical and systematic uncertainties. The preliminary branching fractions based on 281 pb$^{-1}$ data are listed in Table \ref{Tab:DBF} and compared to the 2004 PDG averages \cite{PDG04}, which does not include our earlier results based on $56$ pb$^{-1}$ data \cite{CLEO:DBF56}, in Fig. \ref{Fig:DBFvsPDG}.

\begin{table}[ht]
  \centering
  \caption{ \it Preliminary $D^0$ and $D^+$ branching fractions with statistical and systematic uncertainties. 
    }
  \vskip 0.1 in
  \begin{tabular}{|l|c|} \hline
    Decay & ${\cal B}(\%)$ \\
    \hline
    \hline
    $D^0 \to K^- \pi^+$ & $3.88 \pm 0.04 \pm 0.09$ \\
    $D^0 \to K^- \pi^+ \pi^0$ & $14.6 \pm 0.1 \pm 0.4$ \\
    $D^0 \to K^- \pi^+ \pi^- \pi^+$ & $8.3 \pm 0.1 \pm 0.3$ \\
    \hline
    $D^+ \to K^- \pi^+ \pi^+$ & $9.2 \pm 0.1 \pm 0.3$ \\
    $D^+ \to K^- \pi^+ \pi^+ \pi^0$ & $6.0 \pm 0.1 \pm 0.2$ \\
    $D^+ \to K^0_S \pi^+$ & $1.55 \pm 0.02 \pm 0.05$ \\
    $D^+ \to K^0_S \pi^+ \pi^0$ & $7.2 \pm 0.1 \pm 0.3$ \\
    $D^+ \to K^0_S \pi^+ \pi^- \pi^+$ & $3.13 \pm 0.05 \pm 0.14$ \\
    $D^+ \to K^- K^+ \pi^+$ & $0.93 \pm 0.02 \pm 0.03$ \\
    \hline
  \end{tabular}
  \label{Tab:DBF}
\end{table}

\begin{figure}
  \centering
  \epsfig{file=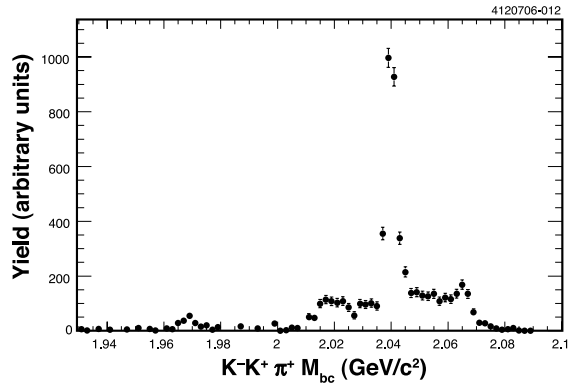,height=2.0in}
  \caption{\it
    $M_{\rm bc}$ distribution for $D^+_s \to K^+K^-\pi^+$ events. The narrow peak at $2.04$ GeV/$c^2$ is due to $D_s$ produced directly, while the broad peak between $2.01-2.07$ GeV/$c^2$ is due to $D_s$ from $D^{\star}_s$ decay.
    \label{Fig:DsMbc} }
\end{figure}

In order to select $D^{\star \pm}_s D_s^{\mp}$ events, we use the beam-constrained mass ($M_{\rm bc} = \sqrt{E_{\rm beam}^2 - (\vec{p}_{D_s})^2}$) and the invariant mass ($M(D_s) = \sqrt{E_{D_s}^2 - (\vec{p}_{D_s})^2}$) of the $D_s$ (or $\bar{D}_s$) candidate and ignore the $\gamma$ or $\pi^0$ resulting from the $D_s^{\star}$ decay. The beam-constrained mass is used as a proxy for the momentum of the $D_s$ candidates (see Fig. \ref{Fig:DsMbc}). We apply a cut on $M_{\rm bc}$ that selects all of the directly-produced $D_s$ and, depending on the decay mode, all or half of the secondary $D_s$. Then the invariant mass of the $D_s$ candidate is used as a primary analysis variable to extract the number of tags. Single tag yields are obtained from fitting the one dimensional $M(D_s)$ distributions, while double tag yields are determined by counting events in the signal regions in the $M(D^+_s)$ vs. $M(D^-_s)$ plane and subtracting backgrounds estimated from sideband regions.

\begin{figure}
  \centering
  \epsfig{file=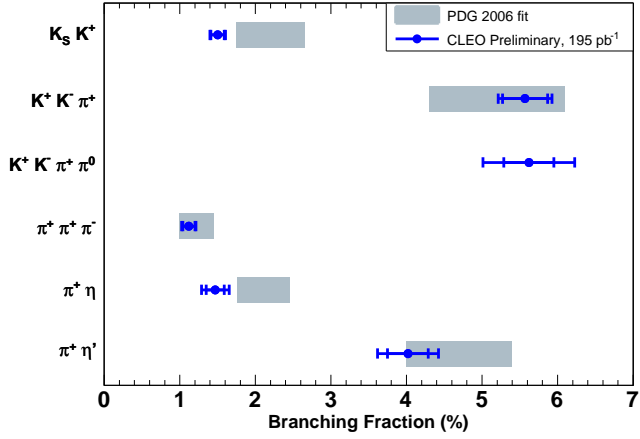,height=2.5in}
  \caption{\it
    Preliminary $D_s$ branching fractions (dots with error bars) compared to the 2006 PDG averages (shaded bars).
    \label{Fig:DsBFvsPDG} }
\end{figure}

For this analysis, we use a binned likelihood hybrid fitter which utilizes Gaussian statistics for single tag modes and Poisson statistics for double tag modes, since the least squares $\chi^2$ fitter used for the $D$ branching fraction measurement is not appropriate for the small signals and backgrounds in the $D_s$ double tag samples. The preliminary branching fractions based on $195$ pb$^{-1}$ data are summarized in Table \ref{Tab:DsBF} and compared to the 2006 PDG averages \cite{PDG06} in Fig. \ref{Fig:DsBFvsPDG}.

\begin{table}[ht]
  \centering
  \caption{ \it Preliminary $D_s$ branching fractions with statistical and systematic uncertainties.
    }
  \vskip 0.1 in
  \begin{tabular}{|l|c|} \hline
    Decay & ${\cal B}(\%)$ \\
    \hline
    \hline
    $D^+ \to K^0_S K^+$ & $1.50 \pm 0.09 \pm 0.05$ \\
    $D^+ \to K^+ K^- \pi^+$ & $5.57 \pm 0.30 \pm 0.19$ \\
    $D^+ \to K^+ K^- \pi^+ \pi^0$ & $5.62 \pm 0.33 \pm 0.51$ \\
    $D^+ \to \pi^+ \pi^+ \pi^-$ & $1.12 \pm 0.08 \pm 0.05$ \\
    $D^+ \to \pi^+ \eta$ & $1.47 \pm 0.12 \pm 0.14$ \\
    $D^+ \to \pi^+ \eta'$ & $4.02 \pm 0.27 \pm 0.30$ \\
    \hline
  \end{tabular}
  \label{Tab:DsBF}
\end{table}

The decay $D^+_s \to \phi \pi^+ \to K^+ K^- \pi^+$, which is one of the largest and easiest to reconstruct, is frequently used as a reference mode to normalize other $D_s$ decays. However, Dalitz plot analysis of this final state by the E687 and FOCUS collaborations has revealed significant signal contribution (from $f^0(980)$ or $a^0(980)$) in the relevant $K^+K^-$ mass region. Because of this extra signal the $\phi \pi^+$ branching fraction might be ill-measured depending on the specific choice of (mass and helicity angle) cuts.
Therefore, we report the partial $D^+_s \to K^+ K^- \pi^+$ branching fraction (${\cal B}_{\Delta M}$) where the mass of the $K^+K^-$ system lies within a $\pm \Delta M$ (in MeV/$c^2$) mass range around the $\phi$ mass ($1019.5$ MeV/$c^2$), which is more useful from experimental point of view than the $\phi \pi^+$ branching fraction. The partial branching fraction with two choices of $\Delta M$ are ${\cal B}_{10} = (1.98 \pm 0.12 \pm 0.09)$\% and ${\cal B}_{20} = (2.25 \pm 0.13 \pm 0.12)$\%.

\section{$D^+$ and $D^+_s$ leptonic decays and decay constants $f_{D_{(s)}}$}

In the Standard Model (SM), purely leptonic decays $D^+_{(s)} \to \ell^+ \nu_{\ell}$ proceed via the annihilation of the constituent quarks into a virtual $W^+$ boson. The decay width is proportional to the decay constant, $f_{D_{(s)}}$, which encapsulates the strong interaction dynamics in the decay:
\begin{equation}
\Gamma(D^+_{(s)} \to \ell^+ \nu) = \frac{G^2_F}{8\pi}m^2M\left(1-\frac{m^2}{M^2}\right)^2 |V_{cd(s)}|^2 f_{D_{(s)}}^2,
\label{Eq:D2ellnuBF}
\end{equation}
$m$ and $M$ are the mass of the charged lepton and the $D_{(s)}$ meson, respectively, $G_F$ is the Fermi coupling constant, $V_{cd}$ ($V_{cs}$) is the relevant quark mixing (CKM) matrix element.

Knowledge of the decay constants is critical for the extraction of CKM matrix elements: e.g. the determination of $V_{td}$ and $V_{ts}$ from measurement of $B \bar{B}$ and $B_s \bar{B_s}$ mixing is limited by the uncertainty in the calculation of $f_B$ and $f_{B_s}$, which currently cannot be measured directly.
Experimental measurement of the $D$ meson decay constants ($f_{D_{(s)}}$) provide an important test of strong interaction theories and validate the most promising calculations involving lattice QCD \cite{LQCD}.

Since the decay width is a function of $m^2$ (helicity suppression), the decay rate to $\tau \nu$ is the largest among the three lepton flavors. Although the $D^+$ ($D^+_s$) decay rate to $\mu \nu$ is a factor of 2.65 (9.72) smaller in the SM, it is easier to measure than the decay to $\tau \nu$ because of the presence of extra neutrino(s) produced by the subsequent decay of the $\tau$. The decay rate to $e \nu$ is suppressed by about five orders of magnitude which is well below the current experimental sensitivity. Any deviation from the SM ratios would be an indication of new physics \cite{NewPhys}.

\begin{figure}[ht]
  \centering
  \epsfig{file=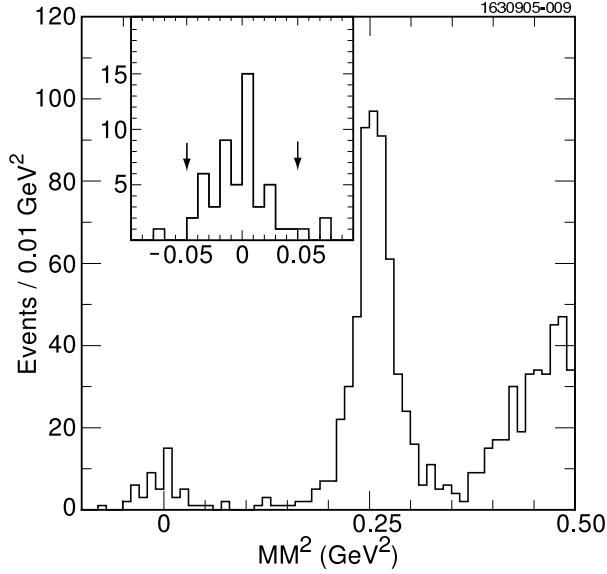,height=3.0in}
  \caption{\it
    The $MM^2$ distribution for $D^+ \to \mu^+ \nu$ candidate events in data. The insert shows the region around zero where the arrows indicate the $\pm 2\sigma$ signal region.
    \label{Fig:D2munu} }
\end{figure}

CLEO has measured the $D^+ \to \mu^+ \nu$ branching fraction in $281$ pb$^{-1}$ data collected at the $\psi(3770)$ \cite{CLEO:D2munu}. We have fully reconstructed the $D^-$ decaying to six hadronic final states ($K^+ \pi^- \pi^-$, $K^+ \pi^- \pi^- \pi^0$, $K^0_S \pi^-$, $K^0_S \pi^- \pi^- \pi^+$, $K^0_S \pi^- \pi^0$, $K^+ K^- \pi^-$), which represent more then 35\% of all $D$ decays. Candidates are selected by requiring $\Delta E$ to be consistent with zero within $2.5 \sigma_{\Delta E}$, and the number of tags in each mode is extracted from a fit to the $M_{\rm bc}$ distribution. The sum of all tags  in the range $-2.5 \sigma_{M_{\rm bc}} < M_{\rm bc} - M_D < 2.0\sigma_{M_{\rm bc}}$ is $158,354 \pm 496$ with a background of $30,677$. In the selected events, we search for a single additional track consistent with a $\mu^+$ and calculate the missing mass squared
\begin{equation}
MM^2 = (E_{\rm beam} - E_{\mu^+})^2 - (-\vec{p}_{D^-} - \vec{p}_{\mu^+})^2,
\end{equation}
where $\vec{p}_{D^-}$ is the three-momentum of the fully reconstructed $D^-$. The $MM^2$ distribution for the data is shown in Fig. \ref{Fig:D2munu}. The peak near zero is mostly due to $D^+ \to \mu^+ \nu$ signal, while the peak at $0.25$ GeV$^2$ is from $D^+ \to \bar{K^0} \pi^+$ decays when a $K_L$ escapes detection.

The signal region within $2 \sigma$ around zero contains $50$ events and the total background is estimated to be $2.8 \pm 0.3 ^{+0.8} _{-0.3}$ events. After background subtraction and efficiency correction, the measured branching fraction is ${\cal B}(D^+ \to \mu^+ \nu) = (4.40 \pm 0.66 ^{+0.09} _{-0.12}) \times 10^{-4}$. The decay constant obtained from Eq. \ref{Eq:D2ellnuBF} using $|V_{cd}| = 0.2238 \pm 0.0029$ and the $D^+$ lifetime ($1.040 \pm 0.007$ ps) is $f_D = (222.6 \pm 16.7 ^{+2.8} _{-3.4})$ MeV. 

We also search for $D^+ \to e^+ \nu$ decay by requiring that the extra track is consistent with an electron and set a $90$\% C.L. upper limit of ${\cal B}(D^+ \to e^+ \nu) < 2.4 \times 10^{-5}$ in the absence of any signal.

\begin{figure}[ht]
  \centering
  \epsfig{file=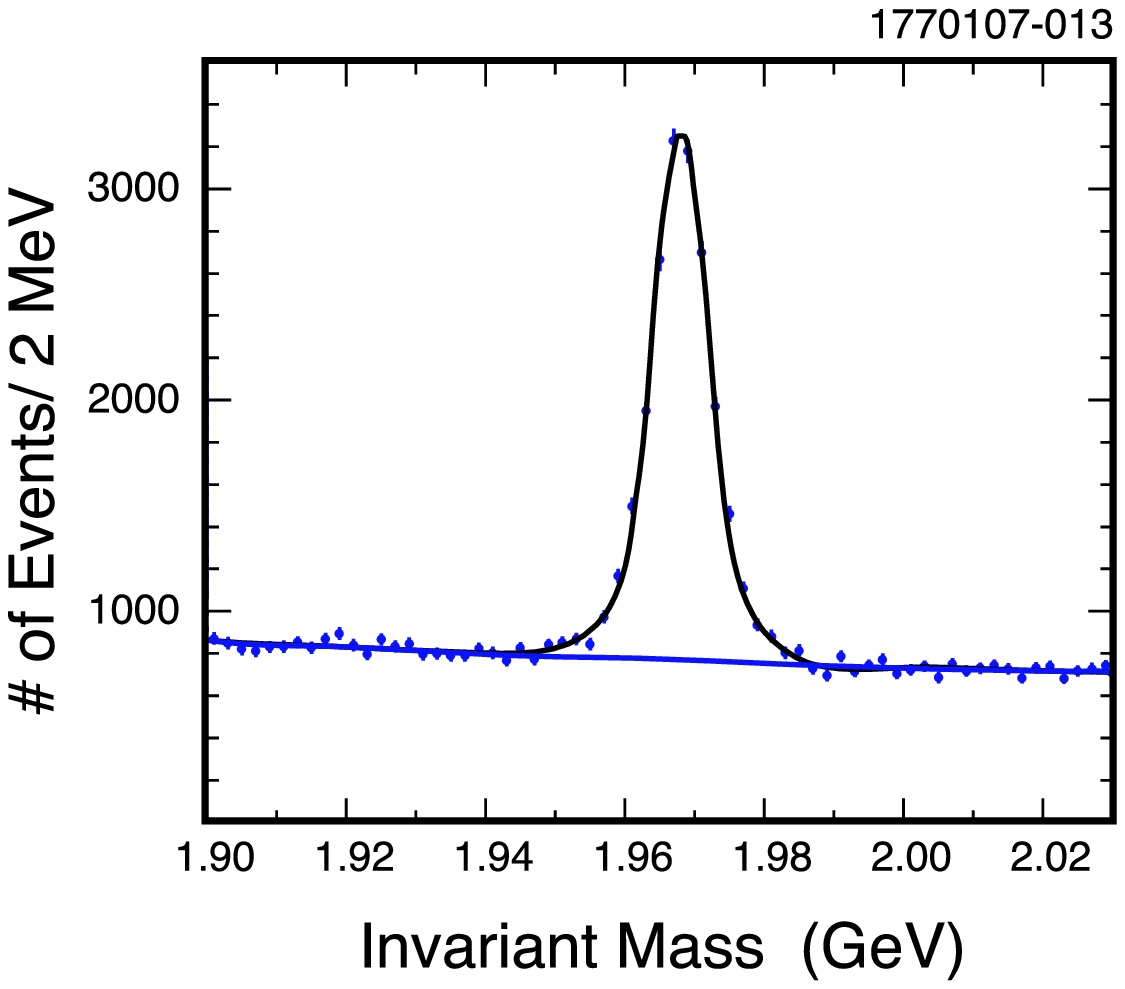,height=2.0in}
  \epsfig{file=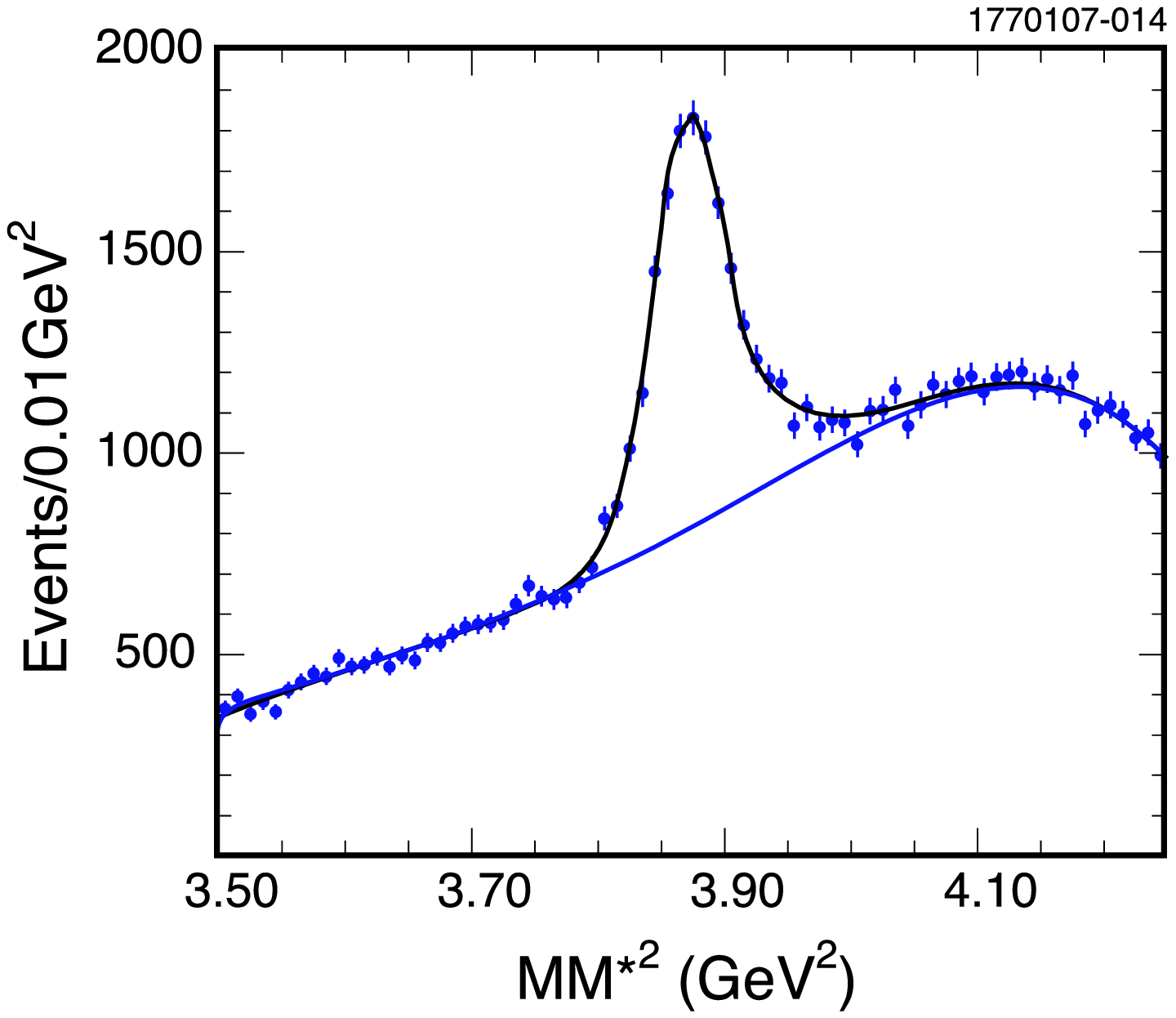,height=2.0in}
  \caption{\it
    Invariant mass of $D^-_s \to K^+ K^- \pi^-$ tags (left) and the recoil mass against the same tag and an additional $\gamma$ (right).
    \label{Fig:Ds2KKpi} }
\end{figure}

The branching fraction of $D^+_s \to \mu^+ \nu$ and $D^+_s \to \tau^+ \nu$ ($\tau^+ \to \pi^+ \bar{\nu}$) is measured in $314$ pb$^{-1}$ data collected at $e^+e^-$ collision energy near $4170$ MeV. We fully reconstruct one $D_s^-$ in eight hadronic decay modes ($K^+K^-\pi^-$, $K^0_S K^-$, $\eta \pi^-$, $\eta' \pi^-$, $\phi \rho^-$, $\pi^+ \pi^- \pi^-$, $K^{\star -} K^{\star 0}$, $\eta \rho^-$). Tags are selected by requiring the beam constrained mass to be $2.015 < M_{\rm bc} < 2.067$ GeV/$c^2$ which is wide enough to accepts both direct as well as secondary $D_s$ from $D^{\star}_s$ decay. The number of tags in each mode is extracted from a fit to the invariant mass distribution of the $D^-_s$ candidates. There is a total of $31,302 \pm 472$ reconstructed tags within $2.5 \sigma$ ($2 \sigma$ for the $\eta \rho^-$ mode) of the $D_s$ mass. In contrast to the hadronic branching fraction measurement, we select a $\gamma$ candidate assumed to be the photon from the $D^{\star}_s \to \gamma D_s$ decay, and then calculate the recoil mass against the $D^-_s$ tag and the $\gamma$:
\begin{equation}
MM^{\star 2} = (E_{\rm cm} - E_{D_s} - E_{\gamma})^2 - (\vec{p}_{\rm cm} - \vec{p}_{D_s} - \vec{p}_{\gamma})^2,
\end{equation}
where $E_{\rm cm}$ ($\vec{p}_{\rm cm}$) the center of mass energy and momentum of the colliding $e^+e^-$ beam. Regardless whether the $D^-_s$ candidate is from the $D^{\star}_s$ decay or not the recoil mass should peak at the $D_s$ mass. We use kinematic constrains to improve the mass resolutions and remove multiple combinations. The recoil mass spectrum of each decay mode is fitted individually to extract the number of $D^{\star}_s D_s$ candidates, which result in a total of $18,645 \pm 426$ events within $2.5 \sigma$ interval around the $D_s$ mass. The invariant mass and recoil mass distributions for $D^-_s \to K^+ K^- \pi^-$ candidates are shown in Fig. \ref{Fig:Ds2KKpi}.

\begin{figure}[ht]
  \centering
  \epsfig{file=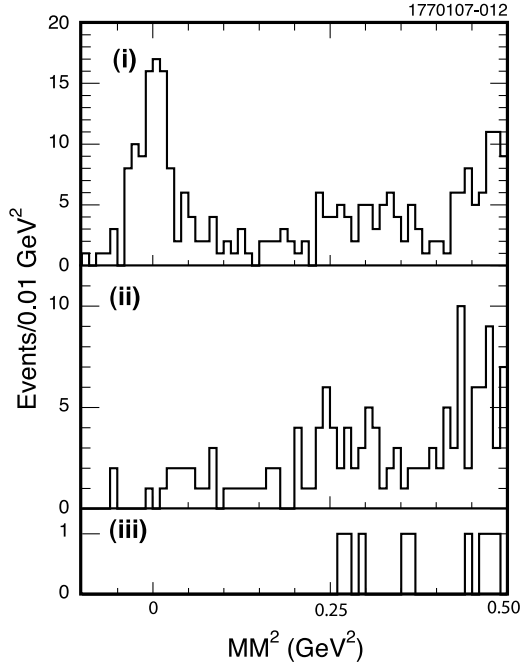,height=3.5in}
  \caption{\it
    The $MM^2$ distribution in data when the additional track is consistent with muon (i), pion (ii), or electron (iii).
    \label{Fig:Ds2ellnu} }
\end{figure}

Then we require a single additional track in the event with opposite charge to the $D_s$ tag and no additional neutral energy cluster with more then 300 MeV. The missing mass is calculated using the energy and momentum of the candidate track ($E_{\mu}$, $\vec{p}_{\mu}$):
\begin{equation}
MM^2 = (E_{\rm cm} - E_{D_s} - E_{\gamma} - E_{\mu})^2 - (\vec{p}_{\rm cm} - \vec{p}_{D_s} - \vec{p}_{\gamma} - \vec{p}_{\mu})^2.
\end{equation}
We consider three cases depending on whether the additional track is consistent with (i) muon (from $D_s \to \mu \nu$), or (ii) pion (from $D_s \to \tau \nu \to \pi \nu \bar{\nu}$), or (iii) electron (from $D_s \to e \nu$). The separation between our muon and pion selection is not complete: the muon selection is $99$\% efficient for muons (with a $60$\% fake rate for pions), while the pion selection accepts 40\% of pions (with a $1$\% fake rate for muons). The $MM^2$ distribution for the three cases is shown on Fig. \ref{Fig:Ds2ellnu}. The peak around zero in (i) is mostly due to $D_s \to \mu \nu$ events. In contrast, $D_s \to \tau \nu \to \pi \nu \bar{\nu}$ events has a longer tail on the positive side due to the extra neutrino. Therefore, we define three signal regions: (A) $-0.05 < MM^2 < 0.05$ GeV$^2$ in (i) for $\mu \nu$ ($92$ events);  (B) $0.05 < MM^2 < 0.20$ GeV$^2$ in (i) and (C) $-0.05 < MM^2 < 0.20$ GeV$^2$ in (ii) for $\pi \nu \bar{\nu}$ ($31$ and $25$ events, respectively). The estimated background from sources other than $D_s \to \mu \nu$ or $\pi \nu \bar{\nu}$ decays is $3.5$, $3.5$, and $3.7$ events, respectively, in the three signal regions.

We calculate three branching fractions: ${\cal B}(D^+_s \to \mu^+ \nu) = (0.594 \pm 0.066 \pm 0.031)\%$ using signal region (A) only; ${\cal B}^{\rm eff}(D^+_s \to \mu^+ \nu) = (0.621 \pm 0.058 \pm 0.032)\%$ from combining all three signal regions (A)+(B)+(C); and ${\cal B}(D^+_s \to \tau^+ \nu) = (8.0 \pm 1.3 \pm 0.4)\%$ from the combined $\tau \nu$ regions (B)+(C). In the first two cases, the $D_s \to \tau \nu$ contribution is subtracted assuming the relative decay rate between $\mu \nu$ and $\tau \nu$ final states is equal to the SM expectation and using ${\cal B}(\tau \to \pi \nu) = (10.90 \pm 0.07)\%$. We also set a $90\%$ C.L. upper limit on ${\cal B}(D^+_s \to e^+ \nu) < 1.3 \times 10^{-4}$.

The decay constant is calculated from the most precise branching fraction (${\cal B}^{\rm eff}$) using Eq. \ref{Eq:D2ellnuBF} with $|V_{cs}| = 0.9730$ and the $D_s$ life time of $(500 \pm 7) \times 10^{-15}$ s: $f_{D_s} = (270 \pm 13 \pm 7)$ MeV. 

We also measure $D^+_s \to \tau^+ \nu$ with a different technique utilizing the $\tau^+ \to e^+ \nu \bar{\nu}$ decay with a total product branching fraction of about $1.3$\%. In this case, we fully reconstruct the $D^-_s$ candidate in the event and require an additional track consistent with an $e^+$ but do not attempt to find the $\gamma$ from the $D^{\star}_s$ decay. Events with additional tracks and more than $400$ MeV total neutral energy in the calorimeter are vetoed (the typical energy of the $\gamma$ or $\pi^0$ from $D^{\star}$ decay is around 150 MeV). After analyzing 195 pb$^{-1}$ subsample of our data we obtain a preliminary branching fraction ${\cal B}(D^+_s \to \tau^+ \nu) = (6.29 \pm 0.78 \pm 0.52)\%$ and decay constant $f_{D_s} = (278 \pm 17 \pm 12)$ MeV.

The weighted average of these two results is $f_{D_s} = (273 \pm 10 \pm 5)$ MeV. Combined with our published $f_D$ value we find a ratio $f_{D_s}/f_D = 1.22 \pm 0.09 \pm 0.03$. The measured decay constants are consistent with most theoretical models. In particular, recent unquenched Lattice QCD calculations \cite{LQCD:Ds} yield $f_{D} = (201 \pm 3 \pm 17)$ MeV, $f_{D_s} = (249 \pm 3 \pm 16)$ MeV, and $f_{D_s}/f_D = 1.24 \pm 0.01 \pm 0.07$.

\section{Measurement of $D^0$ mass}

Precise knowledge of the $D^0$ mass is not only important for its own sake but it can also help with the interpretation of the $X(3872)$ state. Because of the proximity of the $X$ mass ($3871.2 \pm 0.5$ MeV/$c^2$) to $M({D^0}) + M({D^{\star 0}})$, one theoretical suggestion is that the $X(3872)$ is a bound state of $D^0$ and $\bar{D}^{\star 0}$ mesons \cite{X3872}. However, it is necessary to measure the $D^0$ mass with better precision than the current PDG average of $1864.1 \pm 1.0$ MeV/$c^2$ \cite{PDG06} in order to reach a firm conclusion.

CLEO has measured the $D^0$ mass \cite{CLEO:D0mass} in $e^+e^- \to \psi(3770) \to D^0 \bar{D^0}$ events using the decay $D^0 \to K^0_S \phi$ followed by $K^0_S \to \pi^+ \pi^-$ and $\phi \to K^+ K^-$. In order to obtain a clean sample of signal events, the $\bar{D^0}$ has been reconstructed using the same tagging technique described in section \ref{Sec:Dhadron}, imposing loose requirements on $\Delta E$ and $M_{\rm bc}$ of the candidates. The $D^0 \to K^0_S \phi$ decay was selected because the final state pions and kaons have small momenta, and therefore the uncertainty in their measurements makes small contribution to the final result. In addition, the mass of the $K^0_S$ candidates can be kinematically constrained to its well known value.

\begin{figure}[ht]
  \centering
  \epsfig{file=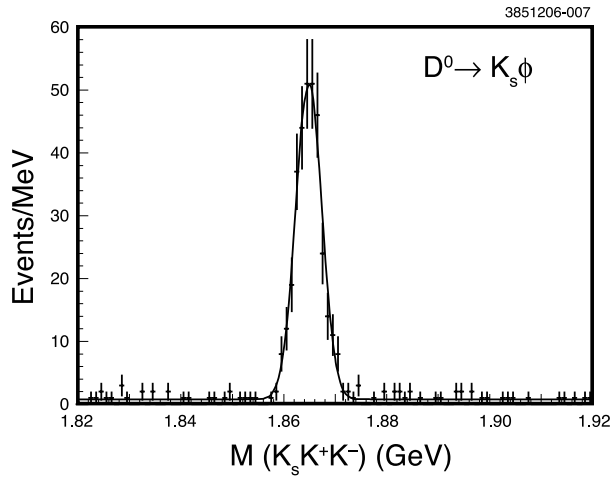,height=2.5in}
  \caption{\it
    The invariant mass of $D^0 \to K^0_s \phi$ candidates in data.
    \label{Fig:D0mass} }
\end{figure}

Pions from the $K^0_S$ are required to originate from a displaced vertex and have a $M(\pi^+\pi^-)$ invariant mass in the range $497.7 \pm 12.0$ MeV/$c^2$ before the mass-constrained kinematic fit. The $\phi$ candidates are accepted with a $M(K^+K^-)$ invariant mass of $1019.46 \pm 15$ MeV/$c^2$. The mass distribution of the $D^0$ candidates in $281$ pb$^{-1}$ data is shown in Fig. \ref{Fig:D0mass}. A likelihood fit using a Gaussian peak and a constant background yields $319\pm18$ $D^0$ events and a $D^0$ mass of $1864.847 \pm 0.150$ MeV/$c^2$ with a mass resolution of $2.52 \pm 0.12$ MeV/$c^2$ (the errors are statistical only). The total systematic error on the mass measurement ($0.095$ MeV/$c^2$) is  dominated by uncertainty in detector calibration, which is studied using the $K_S$ mass in inclusive $D \to K^0_S X$ decays and the $\psi(2S)$ mass in exclusive $\psi(2S) \to \pi^+ \pi^- J/\psi (J/\psi \to \mu^+ \mu^-)$ events.

Our final $D^0$ mass with statistical and systematic uncertainties is
\begin{equation}
M(D^0) = 1864.847 \pm 0.150 \pm 0.095 \ \ {\rm MeV}/c^2.
\end{equation}
This gives $M(D^0 \bar{D}^{\star 0}) = 2M(D^0) + \Delta M_{D^{\star 0} - D^0} = 3871.81 \pm 0.36$ MeV/$c^2$, and leads to a binding energy of the X(3872) as a $D^0 \bar{D}^{\star 0}$ molecule: $\Delta E_b = M(D^0 \bar{D}^{\star 0}) - M(X) = +0.6 \pm 0.6$ MeV/$c^2$. The error in the binding energy is now dominated by the uncertainty in the mass of the $X(3872)$.

\section{Study of $\chi_{cJ} \to h^+ h^- h^0$ decays}

In contrast to the $1^{--}$ members of the charmonium states ($J/\psi$, $\psi(2S)$), the decays of the $\chi_{cJ}$ ($J=0,1,2$) states are not well studied. The different decay mechanism of these states (dominated by annihilation into two (virtual) gluons and contribution from the color-octet mechanism) might provide complimentary information on light hadron spectroscopy and possible glueball dynamics \cite{chi_cJ}.

\begin{figure}[ht]
  \centering
  \epsfig{file=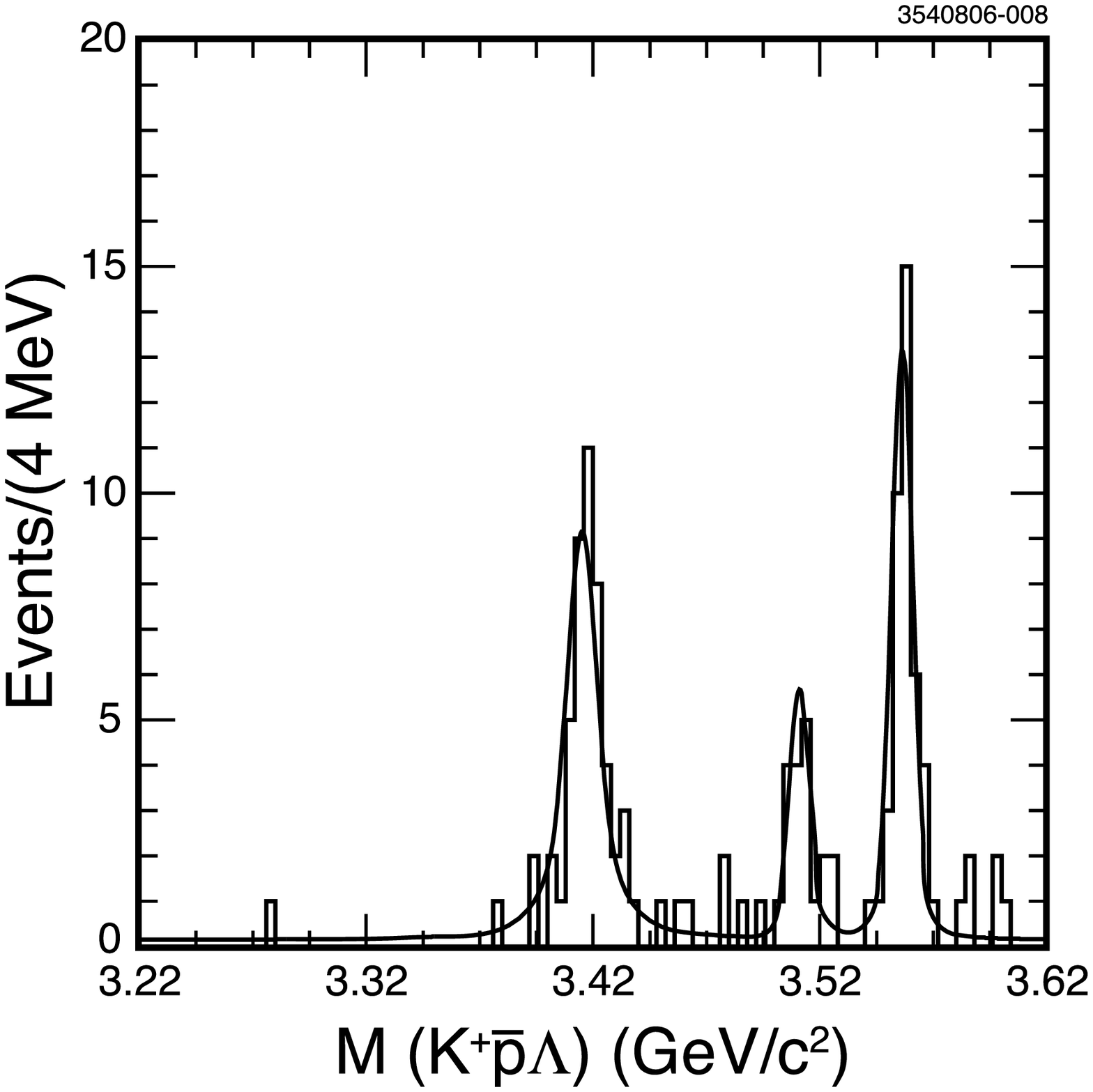,height=2.0in}
  \epsfig{file=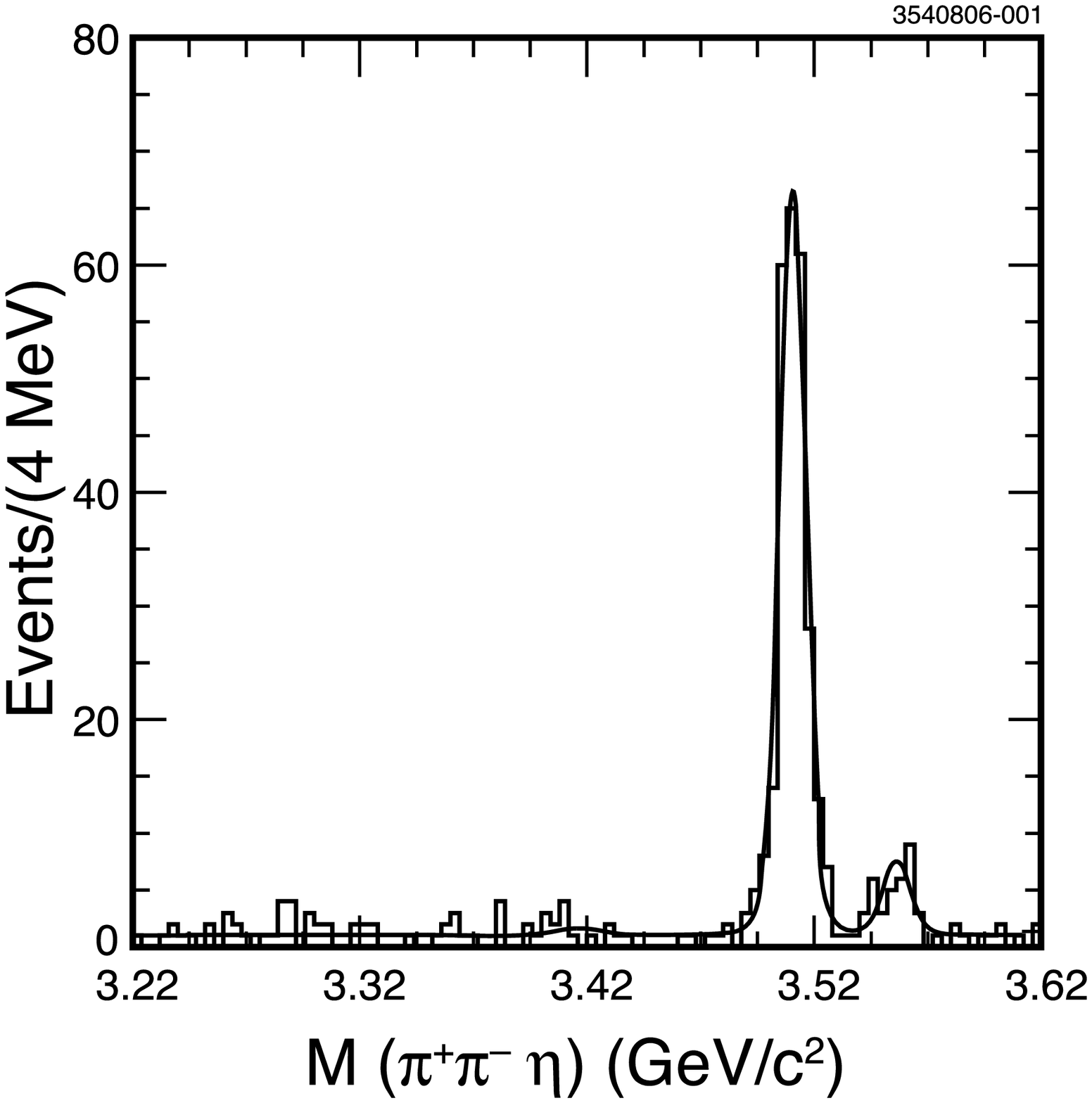,height=2.0in}
  \caption{\it
    The invariant mass distribution for $\chi_{cJ} \to K^+ \bar{p} \Lambda$ (left) and $\chi_{cJ} \to \pi^+ \pi^- \eta$ (right) candidate events in data.
    \label{Fig:KantipLambda} }
\end{figure}

At CLEO, the $\chi_{cJ}$ states are produced in radiative decays of the $\psi(2S)$ and we study their decays to eight selected three-body hadronic modes: $\pi^+\pi^-\eta$, $K^+K^-\eta$, $p\bar{p}\eta$, $\pi^+\pi^-\eta'$, $K^+K^-\pi^0$, $p\bar{p}\pi^0$, $\pi^+K^-K_S$, and $K^+ \bar{p} \Lambda$. We have measured branching fractions or set upper limits for the first time in most cases using about $3$ million $\psi(2S)$ decays \cite{CLEO:chi_cJ}. As an example, Fig. \ref{Fig:KantipLambda} illustrates the invariant mass distribution for two of the hadronic final states.

\begin{figure}[ht]
  \centering
  \epsfig{file=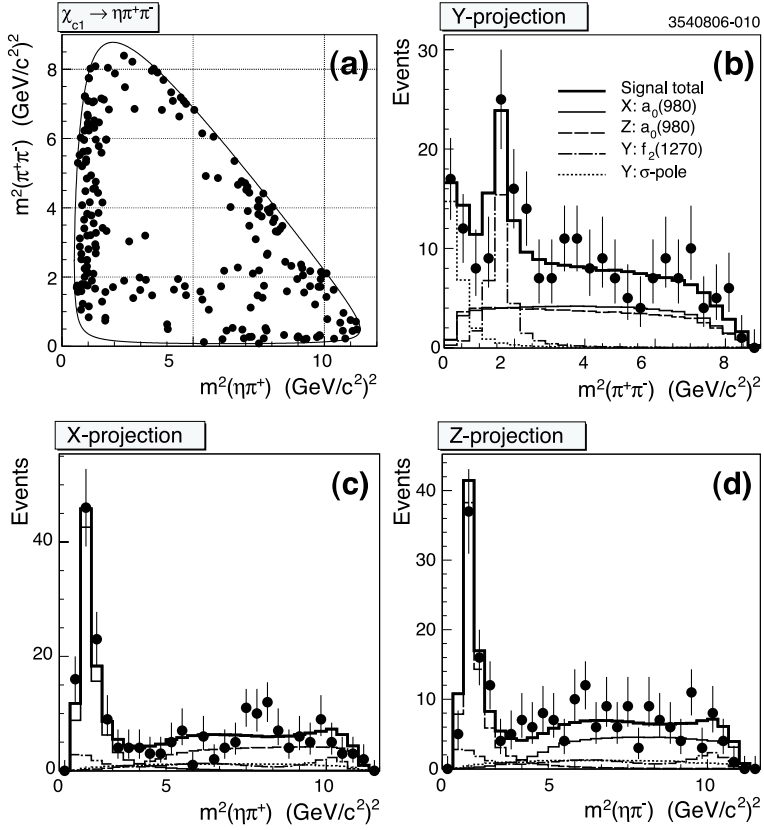,height=4.4in}
  \caption{\it
    Dalitz plot and projections of $\chi_{c1} \to \pi^+\pi^-\eta$ decay.
    \label{Fig:Dalitz} }
\end{figure}

We perform a Dalitz-plot analysis of the decays with the highest statistics, $\chi_{c1} \to \pi^+\pi^-\eta$ (228 events), $K^+K^-\pi^0$ (137 events), and $\pi^+K^-K^0_S$ (234 events) in order to study the two-body resonant substructure. We use a simplified model with non-interfering resonances, which is adequate to show the largest contributions in our small sample. Fig. \ref{Fig:Dalitz} shows the Dalitz plot and three projections for $\chi_{c1} \to \pi^+\pi^-\eta$ and the result of the fit.
There are clear contributions from $a_0(980)^{\pm} \pi^{\mp}$ and $f_2(1270)\eta$ intermediate states, and a significant accumulation of events at low $\pi^+\pi^-$ mass which can be described by an S-wave ($\sigma$) resonance. This mode might offer the best measurement of the $a_0(980)$ parameters with higher statistics. The decays $\chi_{c1} \to K^+K^-\pi^0$ and $\pi^+K^-K^0_S$ are analyzed simultaneously taking advantage of isospin symmetry. We observe contributions from $K^{\star}(892)K$, $K^{\star}(1430)K$ , $a_0(980)\pi$ intermediate states. It is not clear whether the $K^{\star}(1430)$ is $K^{\star}_0$ or $K^{\star}_2$, and other $K\pi$ and $KK$ resonances can contribute. Addition of $\kappa K$ or non-resonant component does not improve the fit and the significance of their contribution remains under 3 standard deviation.

More data is required to do a complete partial-wave analysis taking into account the $\chi_{c1}$ polarization properly and including interference among the resonances.

\section{Summary}

I have reported mostly preliminary results for hadronic and purely leptonic decays of $D$ and $D_s$ mesons from the CLEO-c experiment. These results represent substantial improvement over previous measurements. The $D^+$ and $D^0$ hadronic branching fractions are limited by systematic uncertainties of up to $3$\%. The $D_s$ hadronic branching fractions are measured with relative uncertainties between $6-12$\%, which are dominated by statistics. The measurement of the decay constants $f_D$ ($f_{D_{(s)}}$) from purely leptonic decays are also statistics limited with a total relative uncertainty of $8$\% ($4$\%).

I have also presented the most precise measurement of the $D^0$ mass, and a study of three-body hadronic decays of the $\chi_{cJ}$ states.

Precision of these and other measurements will improve in the near future with more data on the way. CLEO-c has already collected an additional 8 times more data on the $\psi(2S)$, and we plan to increase the $D\bar{D}$ and $D^{\star \pm}_s D^{\mp}_s$ data samples by a factor of 2-3 before data taking ends in April 2008.

\section{Acknowledgements}
I would like to thank the conference organizers for the invitation and warm hospitality, and acknowledge my colleagues at CLEO and CESR for their hard work in achieving the results presented in this report. This research was supported by the US National Science Foundation.

\end{document}